\documentstyle[12pt,epsf,epsfig]{article}
\def\be{\begin{equation}}
\def\ee{\end{equation}}
\def\ba{\begin{eqnarray}}
\def\ea{\end{eqnarray}}
\def\la{~\mbox{\raisebox{-.6ex}{$\stackrel{<}{\sim}$}}~}
\def\ga{~\mbox{\raisebox{-.6ex}{$\stackrel{>}{\sim}$}}~}
\def\bq{\begin{quote}}
\def\eq{\end{quote}}

 at 10truept

\newcommand{\beq}{\begin{equation}}
\newcommand{\eeq}{\end{equation}}
\newcommand{\beqa}{\begin{eqnarray}}
\newcommand{\eeqa}{\end{eqnarray}}

\def\la{~\mbox{\raisebox{-.6ex}{$\stackrel{<}{\sim}$}}~}
\def\ga{~\mbox{\raisebox{-.6ex}{$\stackrel{>}{\sim}$}}~}

\def\ltap{\ \raise.3ex\hbox{$<$\kern-.75em\lower1ex\hbox{$\sim$}}\ }
\def\gtap{\ \raise.3ex\hbox{$>$\kern-.75em\lower1ex\hbox{$\sim$}}\ }
\def\gl{\ \raise.5ex\hbox{$>$}\kern-.8em\lower.5ex\hbox{$<$}\ }
\def\roughly#1{\raise.3ex\hbox{$#1$\kern-.75em\lower1ex\hbox{$\sim$}}}

\begin{document}

\thispagestyle{empty}
\vspace*{1cm}
\begin{center}
{\Large \bf N-flationary magnetic fields}\\
\vspace*{1.5cm} {\large Mohamed M. Anber\footnote{\tt
manber@physics.umass.edu} and
Lorenzo Sorbo\footnote{\tt sorbo@physics.umass.edu}}\\
\vspace{.15cm} {\em Department of Physics,
University of Massachusetts, Amherst, MA 01003}\\
\vspace{.15cm} \vspace{1.5cm} ABSTRACT 
\end{center}

There is increasing interest in the role played by pseudo Nambu-Goldstone bosons (pNGBs) in the construction of string--inspired models of inflation. In these models the inflaton is expected to be coupled to gauge fields, and will lead to the generation of magnetic fields that can be of cosmological interest. We study the production of such fields mainly focusing on the model of N-flation, where the collective effect of several pNGBs drives inflation. Because the produced fields are maximally helical, inverse cascade processes in the primordial plasma increase significantly their coherence length. We discuss under what conditions inflation driven by pNGBs can account for the observed cosmological magnetic fields. A constraint on the parameters of this class of inflationary scenarios is also derived by requiring that the magnetic field does not backreact on the inflating background.

\vskip1.5cm

\section{Introduction}

Observations strongly support inflation as the leading candidate for the description of the first moments of life of the Universe. However, we are still missing a model of inflation that is realistic from the point of view of particle physics. In particular, the achievement of stability of the inflaton potential under radiative corrections is often problematic. As usual, our best hope to find radiative stability lies in symmetry: the required flatness of the inflaton potential can be protected against dangerous radiative corrections if we rely on a (broken) shift symmetry. This situation is realized in models of {\em natural inflation}~\cite{Freese:1990rb}, where the inflaton is a pseudo Nambu-Goldstone boson (pNGB). 

Many of the recent efforts aiming at the construction of models of inflation in string theory contain some variant of natural inflation as a crucial ingredient. For instance, in the model of~\cite{Blanco-Pillado:2004ns} inflation proceeds mostly along an axionic direction. In this context, one of the main difficulties is that sufficient inflation requires an axion constant larger than the Planck mass. This condition, that can be consistently satisfied in field--theoretical constructions~\cite{naturalFT}, seems to be very difficult (if not impossible) to achieve in string theory~\cite{largef}. For this reason, string--inspired models of natural inflation have to resort to more complicated constructions, often invoking more than a single pNBG, such as in the two-axion potentials of~\cite{Kim:2004rp,Blanco-Pillado:2006he}.
More recently, Dimopoulos et al.~\cite{Dimopoulos:2005ac} have shown that inflation can find a natural realization in string theory, if several hundreds of pNGBs roll along their potential. Indeed, even if a single pNGB with an axion constant smaller than $M_P$ cannot sustain inflation for a long time, the collective effect of several such fields ({\em N-flation}) has this possibility. This mechanism -- that could also be responsible for the current epoch of accelerated expansion~\cite{Kaloper:2005aj} --  can occur quite naturally in string theory, where hundreds of axion modes can exist. 

Axions are coupled to gauge fields, so natural inflation (and N-flation in particular) is generically expected to lead to the production of magnetic fields at cosmological scales. 

Magnetic fields are present throughout the whole Universe. Fields with a  strength of $\sim \mu $G and coherent over very large scales ($\sim$kpc--Mpc) have been observed using a variety of techniques. In our Galaxy, the existence of fields of strength 3$-$4$ \mu$G is inferred by observing Zeeman splitting of spectral lines,  synchrotron emission, and Faraday rotation \cite{Kronberg:1993vk}. In addition, magnetic fields coherent over tens of kpc and of strength 1$-$10 $\mu$G have been observed in clusters \cite{ref2} and may play an important role in cluster dynamics . 

The origin of these magnetic fields remains mysterious. A popular mechanism that could be responsible for the large magnetic fields observed is the dynamo mechanism. This mechanism, that sets in after galaxy formation, converts the kinetic energy of a conducting fluid into magnetic energy. However, the dynamo is only a means of amplification, and a seed field of primordial origin is still required. Due to the current evidences in favor of an accelerated expansion of the Universe,  this seed field can be as small as $10^{-30}$~G on a length of $\sim 10$~kpc, significantly weaker than previously thought~\cite{Davis:1999bt} (for comprehensive reviews of magnetic fields in the early Universe see for instance~\cite{reviews}). 

The possibility that a rolling pNGB in the early Universe could lead to the generation of the required seed field was first envisaged by Turner and Widrow~\cite{Turner:1987bw}, and subsequently examined in detail by Garretson, Field and Carroll~\cite{Garretson:1992vt}, with negative conclusions. First, in the most natural setting where the pNGB is the (slowly rolling) inflaton, the smallness of its time derivative appeared to make the mechanism inefficient. More generally, \cite{Garretson:1992vt} has shown that the mechanism acts only when the relevant wavelength is inside the horizon. As the mode exits the horizon, the magnetic field strength starts redshifting away. If we require that the redshifted magnetic field is sufficiently strong to initiate the dynamo, then the energy initially stored in cosmologically interesting modes should have been enormous, much larger than the energy stored in the inflaton. 
This is in contrast with the assumption that the energy in magnetic modes is negligible with respect to the background inflaton energy.

Despite these difficulties, the increasing importance of axions in string theoretical realizations of inflation makes it worth re--examining the issue of the production of  cosmological magnetic fields in this context. Developments in model building  together with new findings about the evolution of magnetic fields in the cosmological plasma can lead to conclusions that are more optimistic than the ones of \cite{Garretson:1992vt}. Even though our main focus will be N-flation, our formulae will also be valid for other scenarios of natural inflation. As we will see, a large number of pNGBs has the potential of effectively increasing the coupling of the inflaton to the magnetic field, thus providing an efficient source even if the pNGBs are rolling slowly. The backreaction problem pointed out in~\cite{Garretson:1992vt} is alleviated by observing that the field produced by this mechanism has maximal helicity. In the last few years, several works have shown that for a maximally helical field an efficient transfer of power from smaller to larger scales (inverse cascade\footnote{While there is a substantial body of literature about inverse cascades of helical fields produced at the electroweak phase transition (see e.g. \cite{inverse,Banerjee:2004df}), we are not aware of analogous works about fields of inflationary origin.}) occurs \cite{inverse,Banerjee:2004df,hind}. When this effect is taken into account, a magnetic field that is initially weak enough not to backreact on the inflating background can inverse cascade to cosmologically interesting scales where its strength is sufficient to start the dynamo.

Our work is organized as follows. In the first section we review the mechanism of N-flation and we derive the relevant equations for the gauge field. Then we study the amplification of the vacuum fluctuations of the magnetic field. In section $4$ we review the argument of~\cite{Garretson:1992vt} and we establish limits on the allowed parameters of our model, while in section $5$ we estimate the intensity of the magnetic field once the inverse cascade process is taken into account. We discuss our results and conclude in section $6$.

\section{Formulation}

We will mainly focus on the model proposed in~\cite{Dimopoulos:2005ac}, that we will quickly review here, and whose mechanism is closely related to that of assisted inflation~\cite{assisted}. We consider a model with $N$ pNGBs. The lagrangian density for this system is given by
\begin{eqnarray}
\label{a1}
{\cal {L}}=-\sqrt{-g}\,\sum _{i=1}^N \left\{\frac{1}{2}\left(\partial\phi_{i}\right)^{2}+\Lambda_{i}^{4}\left[1+\cos(\phi_{i}/f_i) \right]\right\}\,\,.
\end{eqnarray}
We then assume (following \cite{Dimopoulos:2005ac}) that the $N$ axion potentials have the same parameters\footnote{This is not a very strong assumption, since, in order to have agreement with observations, the axion masses have to be very densely packed~\cite{Kim:2006ys}.}, i.e. $\Lambda_{i}=\lambda$, $f_{i}=f$. In a realistic model axions can roll down either side of the cosine potential. In order to take into proper account this fact without complicating the analysis we will assume that the axions have the same absolute value of the initial conditions, but can have either sign i.e $\phi_{i}=s_{i}\tilde \phi$ where $s_{i}=+1,\,-1$. We expect that the main features of the scenario will not change significantly when the above assumptions are dropped\footnote{See e.g.~\cite{Alabidi:2005qi} for an analysis of the predictions of theories with several inflatons.}.

The simplified lagrangian reads
\begin{eqnarray}
\label{a3}
{\cal {L}}=-\sqrt{-g} \left\{\frac{N}{2}\left(\partial\tilde\phi\right)^{2}+N\, \Lambda^{4}\left[1+\cos(\tilde\phi/f) \right] \right\}\mbox{ ,}
\end{eqnarray}
and a canonically normalized axion can be obtained by defining a new field $\Phi=\sqrt N\tilde\phi$, from which we obtain
\begin{eqnarray}
\label{a4}
{\cal {L}}=-\sqrt{-g} \left\{ \frac{1}{2}\left(\partial\Phi\right)^{2}+N\Lambda^{4}\left[1+\cos\left(\frac{\Phi}{\sqrt N f} \right) \right]  \right\}\mbox{ .}
\end{eqnarray}

Therefore, a theory with $N$ axions with axion constant $f$ turns out to be equivalent to the theory of $1$ axion with constant $\sqrt{N}\,f$. Sufficient inflation requires an axion constant larger than about $3\,M_P$~\cite{Savage:2006tr}, while string theory appears to tolerate only $f<M_P$~\cite{largef}. Assuming a sufficiently large value of $N$, it is possible to effectively raise the value of $f$, thus obtaining sufficient inflation without contradicting any theoretical bounds.

Let us now couple the axions to a $U(1)$ gauge field. We parametrize the coupling of the $i$-th axion to the electromagnetic field by $\alpha_i/M_P$. We expect generically to have $\alpha_i\simeq 1$, that represents also the worst case scenario (as usual, we do not expect weaker-than-gravitational couplings to appear in a theory coupled to gravity). 

Based on the considerations above, we assume the coupling of the $N$ axions to the gauge field  to have the following form
\begin{eqnarray}
\label{truecoupl}
{\cal {V}}_{\phi\gamma}=\sum _{i=1}^N \frac{\alpha_i}{4M_{P}}\phi{_i} F_{\mu\nu}\tilde F^{\mu\nu} \mbox{ ,}
\end{eqnarray}
where $F_{\mu\nu}$ is the electromagnetic field strength tensor and $\tilde F^{\mu\nu}=\frac{1}{2}\epsilon^{\mu\nu\rho\sigma}F_{\rho\sigma}$ is the dual tensor. In what follows we will assume for simplicity that all the $\alpha_i$ have the same value, $\alpha_i=\alpha$.

Using the above definition of $\tilde\phi$ and the canonically normalized effective field $\Phi$ this coupling takes the form
\begin{eqnarray}
\label{effectcoupl}
{\cal {V}}_{\Phi\gamma}=\frac{\alpha\gamma\sqrt{N}}{4M_{P}}\Phi F_{\mu\nu}\tilde F^{\mu\nu} \mbox{ .}
\end{eqnarray}
where $\gamma\equiv\sum_{i=1}^{N} s_{i}/N$, $0\leq\mid \gamma \mid\leq 1$
\footnote{In section $6$ we will discuss the possible values that the parameter $\gamma$ is expected to take.}.

The equations of motion for this system are given by
\begin{eqnarray}
\label{a6}
&&\nabla^{\mu}\nabla_{\mu}\Phi-\frac{dV(\Phi)}{d\Phi}=\frac{\alpha\,\gamma\sqrt N}{4M_{P}}F_{\mu\nu}\tilde F^{\mu\nu} \mbox{ ,}\\
\label{a7}
&&\nabla_{\mu}F^{\mu\nu}=-\frac{\alpha\,\gamma\sqrt N}{M_{P}}\left(\nabla_{\mu}\Phi\right)\tilde F^{\mu\nu}\mbox{ ,}\\
&&\label{a8}\nabla_{\mu}\tilde F^{\mu\nu}=0 \mbox{ .} 
\end{eqnarray}
where $\nabla^{\mu}$ denotes the covariant derivative.

The electromagnetic field strength tensor is given in the conformal FRW metric by
\begin{equation}
\label{a9}
F^{\mu\nu}=a^{-2}\pmatrix{0&E_{x}&E_{y}&E_{z}\cr
	-E_{x}&0&B_{z}&-B_{y}\cr
	-E_{y}&-B_{z}&0&B_{x}\cr
      -E_{z}&B_{y}&-B_{x}&0\cr} \mbox{. }
\end{equation}
Plugging eq.~(\ref{a9}) into eq.~(\ref{a6}), one has for the equation of motion of $\Phi$
\begin{equation}
\label{a11}
\frac{\partial^{2}\Phi}{\partial \tau^{2}}+2aH\frac{\partial \Phi}{\partial \tau}-\nabla^{2}\Phi+a^{2}\frac{dV(\Phi)}{d\Phi}=\frac{\alpha\,\gamma\sqrt{N}}{M_{P}}a^{2}\vec E \cdot \vec B \mbox{ ,}
\end{equation}
where $ H$ is the Hubble parameter $ H=a'(\tau)/a^{2}(\tau)$, and where the prime denotes differentiation with respect to the conformal time $\tau$.
Similarly, the equations of motion of $F^{\mu\nu}$ (\ref{a7}) become
\begin{eqnarray}
\label{a12}
\frac{\partial}{\partial \tau}\left(a^{2}\vec E \right)-\nabla\times\left(a^{2}\vec B\right)=-\frac{\alpha\,\gamma\sqrt N}{M_{P}}\frac{\partial \Phi}{\partial \tau}\left(a^{2}\vec B\right)-\frac{\alpha\,\gamma\sqrt N}{M_{P}}\left(\vec\nabla\Phi\right)\times\left(a^{2}\vec E\right)\mbox{ ,}
\end{eqnarray} 
and 
\begin{equation}
\label{a13}
\vec\nabla \cdot\vec E=-\frac{\alpha\,\gamma\sqrt N}{M_{P}}\left(\vec\nabla \Phi\right)\cdot \vec B \mbox{ .}
\end{equation}
In addition, the Bianchi identity (\ref{a8}) becomes
\begin{equation}
\label{a14}
\frac{\partial}{\partial \tau}\left(a^{2}\vec B\right)+\nabla\times\left(a^{2}\vec E\right)=0\mbox{ ,} 
\end{equation}
along with
\begin{equation}
\nabla \cdot \vec B=0 \mbox{ .}
\end{equation}

Under the assumption of homogeneity of the inflaton $\Phi$ we can drop all terms $\nabla \Phi$. As we will be considering the limit of weak electromagnetic field, the term on the right hand side of eq.~(\ref{a11}) will be neglected in what follows. In  section $4$ we will discuss the regime of validity of this approximation. Taking the $curl$ of eq.~(\ref{a12}) and using eq.~(\ref{a14}) to eliminate $\vec E$ one obtains
\begin{equation}
\label{a15}
\left(\frac{\partial^{2}}{\partial \tau^{2}}-\nabla^{2}-\frac{\alpha\,\gamma\sqrt N}{M_{P}}\frac{d\Phi}{d\tau}\nabla\times \right)\left(a^{2}\vec B(\tau,\vec x)\right)=0 \mbox{ .}
\end{equation} 
We define $\vec F(\tau,\vec x)=a^{2}\vec B(\tau,\vec x)$, and then we take the Fourier transform of $\vec F$ defined by
\begin{equation}
\vec F(\tau,\vec k)=\frac{1}{\left(2\pi\right)^{3/2}}\int  e^{-i\vec k.\vec x}\vec F(\tau,\vec x) \:d^{3}x \mbox{ ,}
\end{equation}
to obtain
\begin{equation}
\label{a16}
\left(\frac{\partial^{2}}{\partial \tau^{2}}+k^{2}-i\,\frac{\alpha \gamma\sqrt N}{M_{P}}\,\frac{d\Phi}{d\tau}\,\vec k\times \right)\vec F(\tau,\vec k)=0 \mbox{ .}
\end{equation} 
Directing $k$ along the $x$ axis and defining $F_{\pm}=\left(F_{y}\pm iF_{z}\right)/\sqrt{2}$, equation (\ref{a16}) becomes
\begin{equation}
\label{a17}
\frac{\partial^{2}F_{\pm}}{\partial \tau^{2}}+\left(k^{2} \pm \frac{\alpha\,\gamma\sqrt N}{M_{P}}\,\frac{d\Phi}{d\tau}\, k \right)F_{\pm}=0 \mbox{ .}
\end{equation} 

As in \cite{Dimopoulos:2005ac}, we assume that $\sqrt{N}\,f\gg M_P$. In this case the observationally interesting epoch of inflation takes place close to the bottom of the cosine potential of the pNGB. We can thus approximate the potential as
\begin{equation}
V(\Phi)=\frac{m^{2}\Phi^{2}}{2}\,\,,\qquad \mbox{ where } m=\frac{\Lambda^{2}}{f} \mbox{ . }
\end{equation}
At leading order in the slow roll parameter $\epsilon=M_{P}^{2}\,V'^{2}/2\,V^{2}$ the solution of the slow roll equations gives 
\begin{eqnarray}
\frac{d\Phi}{d\tau}\simeq\sqrt{2\epsilon}\:M_{P}\left(-\tau\right)^{-1} \mbox{ ,}
\end{eqnarray}
where the conformal time $\tau$ is related to the scale factor by $a\left(\tau\right)\simeq (-H\,\tau)^{-1}$, and we have assumed that inflation ends at $\tau=-1/H$, where $a=1$.

\section{ Magnetic field production}

The generation of the magnetic field is due to the amplification of quantum fluctuations in the presence of the time varying background provided by the slowly rolling pNGBs\footnote{The possibility of generating a magnetic field of cosmological interest during the stage of coherent oscillations of a pNGB was studied in~\cite{ale}.}. To study this process, we promote the classical field $\vec F(\tau,\vec x)$ to a quantum mechanical operator in the Heisenberg representation:
\begin{equation}
\label{d1}
\hat F_{i}(\tau,\vec x)=\int \frac{d^{3}k}{(2\pi)^{3/2}}\left[\hat a_{k,i}F_{i}(\tau,\,k)e^{i\vec k \vec x}+\hat a^{\dagger}_{k,i}F^{*}_{i}(\tau,k)e^{-i\vec k \vec x}  \right]\mbox{ , }
\end{equation}
where $\hat a_{k,i}$ and $\hat a^{\dagger}_{k,i}$ satisfy the commutation relations $\left[\hat a_{k,i},\hat a^{\dagger}_{k',j}\right]=\delta(\vec k-\vec k')\delta_{ij}$ , $\left[\hat a_{k,i},\hat a_{k',j}\right]=0$ along with $a_{k,i}\vert 0\rangle_{I}=0 $, where $\vert 0 \rangle_{I}$ is the initial vacuum state. The functions $F_{i}(\tau,k)$ satisfy equation (\ref{a16}) and are normalized to give the vacuum solution at $k\tau \rightarrow -\infty$ 
\begin{equation}
\label{d2}
F_{i}(\tau,k)=\sqrt{\frac{k}{2}}e^{-ik\tau} \mbox{. }
\end{equation}

Depending on the sign of $\alpha\,\gamma\,{d\Phi}/{d\tau}$, one of the two solutions $F_{+}$ or $F_{-}$ in (\ref{a17}) will develop an instability which will grow during inflation. In the following analysis we will denote by $F$ the growing solution. 
The general equation for $F$ can be written as
\begin{equation}
\label{d3}
\frac{d^{2}F(\tau,\,\vec k)}{d\tau^{2}}+\left[k^{2}+2k\,\frac{\xi}{\tau} \right]F(\tau,\,\vec k)=0\mbox{ ,}
\end{equation}
where 
\begin{equation}
\xi\equiv\left\vert\alpha\gamma\right\vert\sqrt{N\epsilon/2 }\,\,.
\end{equation} 
As we shall see, we will be interested in the case where $\xi={\cal {O}}\left(1\right)$ or larger. 

In terms of the dimensionless variable $-k\tau$, eq.~(\ref{d3}) is the Coulomb wave equation\footnote{The general Coulomb equation takes the form $\frac{d^{2}F}{d\rho^{2}}+\left[1-\frac{2\xi}{\rho}-\frac{L(L+1)}{\rho^{2}} \right]F=0 $ , where $L=0,1,2,...$.} with $L=0$. The solution of this equation that reduces to positive frequency for $k\tau\rightarrow -\infty$ is
\begin{equation}
\label{d5}
F(\tau,\,\vec k)=\sqrt{\frac{k}{2}}\left[i\,F_{0}(\xi,\,-k\tau)+G_{0}(\xi,\,-k\tau)\right] \mbox{ ,}
\end{equation}
where $F_0$ and $G_0$ are respectively the regular and irregular Coulomb wave functions with index $0$~\cite{Stegun}. At early times, the above solution has the asymptotic behavior
\begin{equation}
\label{d6}
F(\tau,\,\vec k)\sim \sqrt{\frac{k}{2}} \exp\left\{-i\left[ k\tau +\xi\, \mbox{ln}(-2\,k\tau)-\sigma_{0} \right] \right\}\quad\mbox{ as } k\tau \rightarrow -\infty \mbox{ , }
\end{equation}
where $\sigma_{0}=\mbox{arg}\Gamma(1+i\xi)$.

As time evolves, the mode gets rapidly amplified: when the second term in brackets in eq.~(\ref{d3}) dominates over the first one, $\vert k\tau\vert\ll 2\xi$, the solution~(\ref{d5}) is well approximated by
\begin{eqnarray}
\label{approx}
F(\tau,\vec k)&\simeq&
 \sqrt{\frac{k}{2}}\left(\frac{k}{2\xi\,aH}\right)^{1/4}e^{-2\,\sqrt{2\xi \,k/aH}+\pi\,\xi} \mbox{ , }
\end{eqnarray}
where we have used $\tau= -1/aH$.

We thus see that the magnetic field gets amplified by a factor $e^{\pi\xi}$. This represents our main result: for moderately large values of $\xi$ we can get an extremely large value of the magnetic field. Inspection of the behavior of the Coulomb wave functions shows that this instability builds when $2\,\xi \simeq \vert k\tau\vert $.

We have now all the quantities needed to estimate the value of the generated magnetic fields. Before doing this, however, we have one more constraint to take into account.

\section{Backreaction}

As recognized in~\cite{Garretson:1992vt}, the amplification of the magnetic field constrains the allowed values of the Hubble parameter. This constraint comes from the requirement that the energy in the magnetic field should not exceed the energy stored in the inflaton.

The total energy density in the magnetic field is given by
\begin{equation}
\label{rhobtotdef}
\rho_M\left(\tau\right)=\frac{1}{2\,a\left(\tau\right)^4}\int_0^{k_c}\,\left\vert F\left(\tau,\,k\right)\right|^2\,\frac{k^2\,dk}{2\,\pi^2}\,\,,
\end{equation}
where $k_c$ is the ultraviolet cutoff of the magnetic modes that get amplified by the rolling axions. Inspection of eq.~(\ref{d3}) gives
\begin{equation}
\label{kc}
k_c\simeq -2\xi/\tau\simeq 2\xi\,Ha\left(\tau\right)\,\,.
\end{equation}

In the regime $k<k_c$, we can use the approximate solution~(\ref{approx}), that gives the following expression for the total energy in the magnetic field
\begin{equation}
\label{rhobtotfin}
\rho_M\left(\tau\right)=\frac{H^4\,e^{2\pi\xi}}{2^{25}\,\pi^2\,\xi^5}\,\int_0^{8\xi}x^8\,e^{-x}\,dx\simeq 10^{-4}\,\frac{H^4\,e^{2\pi\xi}}{\xi^5}\,\,,
\end{equation}
where in the last expression we have assumed $\xi\ga 1$.

By requiring $\rho_M\la \rho_I=3\,M_P^2H^2$ we get an upper bound on the inflationary Hubble parameter 
\begin{equation}\label{constr}
H\la 150\,\xi^{5/2}e^{-\pi\xi}\,M_P\,\,.
\end{equation}

If we insist on COBE normalization ($H\simeq 10^{13}$~GeV), this implies $\xi\la 7$. Of course, we can also give up the COBE constraint by invoking the existence of one or more curvatons~\cite{kostas} (after all, in a scenario with hundreds of light fields, this is not such an unreasonable expectation). Requiring that inflation takes place at least at the TeV scale ($H\simeq 10^{-3}$~eV) gives the constraint\footnote{It would be interesting to study the behavior of the system when the backreaction cannot be neglected.} $\xi\la 25$.

We also note that in general the axions will be coupled to more than a single $U(1)$ gauge field. In grand unified theories, for instance, few dozens of vector degrees of freedom appear in the same multiplet and will be excited by the slowly rolling axions. The hidden sector can contain even larger gauge groups. As a consequence, the total energy in the excited modes $\rho_M$ should be multiplied by the number of vectors that transform under the same group, leading to stronger constraints on the parameter $\xi$.

If we impose COBE normalization and require that the subsequent evolution of the magnetic field is just given by conservation of the magnetic flux $B\propto B_{\mathrm {initial}}/a^2$, the ensuing field will be too weak to be able to seed the dynamo mechanism~\cite{Garretson:1992vt}. However, the field generated by rolling axions has {\em maximal helicity}, a property that affects significantly its subsequent evolution, as we will see in the next section.

\section{Estimating the magnetic field}

Due to the high conductivity and turbulence of the primordial plasma, magnetic fields will evolve not only conserving magnetic flux but also magnetic helicity $\int d^{3}x \: \vec B \cdot \vec A$, where $\vec A$ is the vector potential. During the last decade, several works have shown that turbulent fluid with non vanishing 
helicity can transfer magnetic energy from small to large scales.  Indeed, a turbulent fluid tends to dissipate the energy in small scale magnetic modes. However, in the regime of large conductivities (that is realized for most of the history of the Universe) helicity has to be conserved. In order to conserve helicity, part of the power that is lost at small scales has to be transferred to large scales, in a phenomenon known as {\em inverse cascade} \cite{inverse}\footnote{The possibility that the effects of turbulence could be relevant for the evolution of magnetic fields produced at inflation was first considered in~\cite{Dimopoulos:1996nq}.}. Since in our model either the positive or the negative helicity field gets amplified, maximal helicity fields are naturally produced. Thus, by properly taking into account the inverse cascade, the backreaction problem can become less severe, allowing to get significant magnetic fields while satisfying COBE normalization. This depends on the ability of this mechanism to produce maximally helical magnetic fields coherent over large scales\footnote{Note also that an existing magnetic field could acquire a helical component due to a QCD axion~\cite{Campanelli:2005ye}.}. We quantify such coherence length by computing the two-point function defined as
\begin{equation}\label{two point function} 
{\cal {G}}_{ij}(\tau,\vec r)=\langle F_{i}\left(\tau,\vec x\right)F_{j}\left(\tau,\vec x+\vec r \right)\rangle =\int \frac{d^{3}k}{\left(2\pi\right)^3}\, e^{-i\vec k \cdot \vec r}F_{i}\left(\tau,\vec k\right)F_{j}^{*}\left(\tau,\vec k\right)\,\,.
\end{equation} 
Taking $\vec r$ to be along the $z$ axis $\vec r=z\,\hat e_{z} $, we can write
 $F_{i}\left(\tau,\vec k \right)=F_{+}\left(\tau,\vec k\right)\epsilon_{i}^{+}+F_{-}\left(\tau,\vec k\right)\epsilon_{i}^{-}$,
 where $\epsilon^{\pm}_{i}=(\epsilon_{i}^{1}\pm \,i\,\epsilon_{i}^{2})/\sqrt 2$ and $\{\vec\epsilon^{\pm}\}$ are the polarization vectors in the $xy$ plane. Thus
\begin{equation}\label{two point expression}
{\cal {G}}_{ij}(\tau,z\,\hat e_{z} )= \int\frac{d^{3}k}{\left(2\pi\right)^3}\, e^{-izk_{z}}\left[\vert F_{+}\vert^{2}\epsilon^{+}_{i}\,\epsilon^{*+}_{j}+\vert F_{-} \vert^{2}\epsilon^{-}_{i}\,\epsilon^{*-}_{j}+F_{+}\,F_{-}^{*}\epsilon^{+}_{i}\,\epsilon^{*-}_{j}+F_{-}\,F_{+}^{*}\epsilon^{-}_{i}\,\epsilon^{*+}_{j}  \right]\,\,.
\end{equation}
Using the relations $\vec\epsilon\,{}^{*}\cdot\vec\epsilon=1$ and  $\vec\epsilon^{*\pm}\cdot\vec\epsilon^{\mp}=0$ we obtain
\begin{equation}\label{correlation of magnetic field}
\sum_{i}{\cal {G}}_{ii}(\tau,z\,\hat e_{z} )=\langle F^{2}(\tau,z)\rangle=\int \frac{d^{3}k}{\left(2\pi\right)^3}\, e^{-izk_{z}}\left[ \vert F_{+}\vert^{2}+\vert F_{-}\vert^{2}\right]\,\,,
\end{equation}
where either $F_{+}$ or $F_{-}$ is nonvanishing. 
Using eq.~(\ref{approx}) we obtain
\begin{eqnarray}
\label{semifinal result}
\langle F^{2}(\tau,z)\rangle &=&\frac{1}{\left(2\pi\right)^2{\sqrt{2\xi\,a\,H}}}\,\int_{0}^{k_{c}} dk\, k^{7/2}\,\frac{\sin kz}{kz}\,e^{-4\sqrt{2\xi\,k/a\,H}+2\pi\,\xi}\,\,.
\end{eqnarray}

We are interested in the spectrum of the magnetic field at the end of inflation, and hence we set $a=1$ in the above expressions. By integrating numerically eq.~(\ref{semifinal result}), it is possible to see that the two-point function at the end of inflation can be well approximated by
\begin{equation}\label{fit}
\langle F^{2}(\tau,L)\rangle\simeq 2\times 10^{-4}\, H^4\,\frac{e^{2\pi\xi}}{\xi^5}\,\frac{1}{\left(1+L/L_c^i\right)^{9/2}}\,\,,
\end{equation}
where the coherence length at the end of inflation $L_c^i$ is given by
\begin{equation}\label{coherence}
L_c^i \simeq 3\,\xi/H\,\,.
\end{equation}

Analytical and numerical studies show that the comoving coherence length for a maximally helical magnetic field in a turbulent plasma grows as $\tau^\alpha$, where the exponent $\alpha$ lies somewhere between $\alpha=1/2$ and $\alpha=2/3$ \cite{inverse,Banerjee:2004df,hind}. In what follows we will assume $\alpha=2/3$. As a consequence, the comoving coherence length $L_c$ increases as $a^{2/3}$ during radiation domination and as $a^{1/3}$ during matter domination. The inverse cascade proceeds during the whole post--inflationary era, until recombination takes place \cite{Banerjee:2004df}. As our field is maximally helical and with a coherence length of few Hubble lengths at the end of inflation, we have several different stages: {\em (i)} as inflation ends, the coherence length is superhorizon, no causal process can affect it and it is just subject to redshift; {\em (ii)} the coherence length enters the horizon while the Universe is dominated by the inflation oscillations that redshift as matter. The conductivity is already very large~\cite{Turner:1987bw}, and we expect inverse cascade to take place. However, since during reheating the temperature does not redshift as $T\propto 1/a$~\cite{Turner:1983he}, the findings of the literature (that are based on conformal transformations from the minkowskian case, thus assuming $T\propto 1/a$) cannot be applied to this regime. In the absence of more reliable results, we will assume that $L_c\propto a^{1/3}$, as in the case of a matter dominated Universe with $T\propto 1/a$. {\em (iii)} After reheating has completed, the Universe is radiation dominated until matter--radiation equality occurs at temperature of $\sim 1$~eV. During this epoch, $L_c\propto a^{2/3}$. {\em (iv)} In the short matter dominated epoch between matter--radiation equality and recombination, we have again $L_c\propto a^{1/3}$. 
Collecting all these factors, we get for the comoving coherence length at recombination $L_c^f$
\begin{equation}
L_c^f=L_c^i\left(\frac{a_{\mathrm{RH}}}{a_{\mathrm{ent}}} \right)^{1/3}\left(\frac{a_{\mathrm{eq}}}{a_{\mathrm{RH}}} \right)^{2/3}\left(\frac{a_{\mathrm{rec}}}{a_{\mathrm{eq}}}\right)^{1/3}\,\,,
\end{equation}
where $a_{\mathrm{ent}}$ is the value of the scale factor when the coherence length enters the horizon. $a_{\mathrm {RH}}$ is related to the reheating temperature $T_{\mathrm {RH}}$ by $H/a_{\mathrm {RH}}^{3/2}=0.33\,g_{*}^{1/2}\,T_{\mathrm {RH}}^{2}/M_P$, where we recall that we have set $a=1$ at the end of inflation and $g_{*}=228.75$ for the MSSM. The physical coherence length can be obtained by multiplying the comoving length by the factor $(a_{0}/a_{\mathrm{end}})=(T_{\mathrm{RH}}/T_{0})\left(M_P\,H/0.33\,g_{*}^{1/2}\,T_{\mathrm{RH}}^{2} \right)^{2/3}$. Using $T_{\mathrm{rec}}=0.26$ eV, $T_{\mathrm {eq}}=0.7$~eV, $H=10^{13}$ GeV, and $T_{0}=2.4\times10^{-4}$ eV we find a coherence length of about one parsec\footnote{We denote with a subscript "phys" the value that would have been taken by the corresponding physical quantities in the absence of the contraction associated to structure formation.}
\begin{equation}
L_{\mathrm{phys}}^{f} \simeq 1.5\,\frac{\xi^{1/3}}{\left(T_{\mathrm {RH}}/10^9\,{\mathrm {GeV}}\right)^{1/9}}\,{\mathrm {pc}}\,\,.
\end{equation}

As far as the intensity of the magnetic field is concerned, conservation of helicity imposes the relation  $B_c^f=B_c^i\left(L_c^i/L_c^f\right)^{1/2}$.  Moreover, numerical analyses~\cite{Banerjee:2004df,hind} show a property of {\em self--similar} evolution of the spectrum of the magnetic field: for lengths larger than the coherence length, the spectral index is left unchanged by the inverse cascade process. Therefore, for $l_c>L_c^f$, the spectrum of the maximally helical magnetic field is given by \cite{Banerjee:2004df}
\begin{equation}\label{comoving magnetic field2}
B_c^f(l_c)=B_c^i\,\left(\frac{L_c^i}{L_c^f}\right)^{1/2}\left(\frac{l_c}{L_c^f} \right)^{-n/2}\,\,,
\end{equation}
where $n=9/2$ is the spectral index for our case. 
The physical field is given as usual by the scaling $B_{\mathrm{phys}}^{f}=B_c^f(l_c)(a_{\mathrm{end}}/a_{0})^2$, so that we obtain finally, for $l_{\mathrm {phys}}>L^f_{\mathrm {phys}}\simeq 1$~pc,
\begin{equation}\label{final}
B\simeq 10^{-33}\,\frac{e^{\pi\,\xi}}{\xi^{17/12}}\,\left(\frac{T_{\mathrm {RH}}}{10^9\mathrm {GeV}}\right)^{11/36}\,\left(\frac{l_{\mathrm {phys}}}{10\,{\mathrm {kpc}}}\right)^{-9/4}\,{\mathrm {G}}\,\,.
\end{equation}

Before analyzing this result, it is worth discussing the strong constraints of~\cite{Caprini:2001nb} on the intensity of magnetic fields of primordial origin. Such constraints emerge from requiring that the magnetic modes do not overproduce gravitational waves in the very early Universe. However, the derivation of~\cite{Caprini:2001nb} assumes that the magnetic field is non helical and that evolves just under the effect of the expansion of the Universe, neglecting the possibility of inverse cascade effects. As a consequence, the analysis of~\cite{Caprini:2001nb} does not apply to our scenario. Further study is needed to find whether analogous bounds apply to helical primordial fields.

\section{Discussion and Conclusion}

We have seen that for a system of $N$ axions with a coupling $\alpha/M_P$ to the photon, the strength of the magnetic field produced turns out to depend only on the combination $\xi=\left\vert \alpha\gamma\right\vert\sqrt{N\epsilon/2}$, where $\gamma=\sum_i s_i/N=\left(N_+-N_-\right)/N$ measures the difference between the number of pNGBs axions rolling to positive values and the ones rolling to negative values. Equation~(\ref{final}) shows that it is possible to get a magnetic field capable of initiating the dynamo for $\xi$ as small as $2$ or so. Now, $\xi$ depends on $\epsilon$, that in turn depends on the time at which the scales of interest have been amplified. Since we have seen that a scale of $\sim 1$ pc today corresponds to the size of the horizon at the end of inflation, we are interested in modes that were amplified roughly $9$ efolds before the end of inflation ($\epsilon\simeq 1/18$). The condition for sufficient amplification is therefore  $\alpha\,\gamma\sqrt{N}\ga   10$. This condition involves only quantities of the order of unity. 

Let us first focus on the role played by the quantity $\gamma$. If the axion potentials are exactly symmetric under $\phi_i\rightarrow -\phi_i$, then for large $N$ $\gamma$ is a random gaussian variable centered at $\gamma=0$ and with variance $1/\sqrt{N}$. Therefore at the $1\sigma$ probability level $\gamma\sqrt{N}=1$ and the system behaves as if there was only one axion. If $\alpha=1$, the condition $\alpha\,\gamma\sqrt{N}>10$ can be realized by mere chance as a 10$\sigma$ effect! To fix ideas, if $N\simeq 600$ as required by the analysis of~\cite{Kim:2006ys}, a sufficiently strong seed field would be achieved if $\sim 420$ pNGBs are rolling in one direction and the remaining $\sim 180$ are rolling in the opposite direction.

However, in general we do not expect the axion potential to be exactly symmetric with respect to the transformation $\phi_i\rightarrow -\phi_i$. Such situation has been for instance envisaged in~\cite{Blanco-Pillado:2004ns} in order to get a densely spaced distribution of vacuum energies (see also~\cite{Arkani-Hamed:2005yv}), where a potential of the form $\Lambda_1^4\cos a\phi+\Lambda_2^4\cos b\phi+\Lambda_3^4\cos \left(a-b\right)\phi$ was considered. In this case, while the potential is still $Z_2$ symmetric around $\phi=0$, there are many other local maxima around which the potential is not $Z_2$ symmetric. In general, inflation is expected to occur on the top of one of these non symmetric maxima. It would be interesting to see if there is any correlation between the details of the axion potential that can lead to a preferred direction in the inflaton path and the possibility of having a sufficiently large value of $\gamma$.

Even in the case of a single pNGB driving inflation (or for $\gamma\la 10/\sqrt{N}$) a moderately large value of $\alpha$ could give the desired result. The precise value of this coupling depends on the details (such as the form of gauge kinetic functions, or the coupling of the axion  to charged fermions) of the model considered and the study of its allowed values is beyond our scope. Let us however mention that, if the "natural" coupling of the axions to the magnetic field is given by $\tilde\alpha/f$ (rather than $\alpha/M_P$) with $\tilde\alpha={\cal {O}}\left(1\right)$ and $f\simeq 0.1 M_P$, then the required ${\cal {O}}\left(10\right)$ enhancement is readily achieved.

We also stress that the requirement that the produced magnetic field does not backreact on the inflaton induces a rather strong bound on the parameter $\xi$, as we have discussed in section $4$. This is a new constraint that has to be satisfied in working models of pNGB inflation.

Let us finally note that the result~(\ref{final}) is just an order of magnitude estimate based on several assumptions (in particular about the evolution of the coherence length of the magnetic field during reheating) and that the accuracy of this result could be improved by a more detailed study. Note also that in this discussion we have assumed $\xi\simeq 2$, that leads to a field of $10^{-30}$~G at $10$~kpc, sufficient to initiate the dynamo according to~\cite{Davis:1999bt}. However, depending on the value of the parameter $\xi$, stronger fields can be generated. The value $\xi=7$, that saturates the bound~(\ref{constr}), leads to fields as strong as $\sim 10^{-25}$~G at $10$~kpc. Elsewhere in the literature one can find requirements stronger than those of~\cite{Davis:1999bt}: for instance, according to the second of refs.~\cite{reviews}, a seed field of $10^{-23}$~G at $1$~Mpc is needed to initiate the dynamo. Such a field can be produced in our scenario if $\xi\simeq 12$. While this assumption violates the bound~(\ref{constr}), it could still be a realistic one if some curvaton--like mechanism is responsible for the generation of the spectrum of primordial perturbations and the COBE requirement $H\simeq 10^{13}$~GeV is relaxed.

To summarize, pNGBs play an increasingly important role in string-motivated models of inflation. For this reason we have reviewed the issue of production of magnetic fields in this context. While in~\cite{Garretson:1992vt} it was shown that such fields cannot be of cosmological interest, the analysis of more recent results in magnetohydrodynamics~\cite{inverse,Banerjee:2004df,hind} can lead to different conclusions. We have found simple formulae that show how cosmologically relevant magnetic fields can be generated in models where some parameters are tuned to be of ${\cal {O}}\left(10\right)$. The same formulae lead to a bound on the parameters of the theory that originates from requiring that the produced magnetic field does not backreact on the inflaton. 

The production of cosmological magnetic fields of sufficient intensity was one of the nice predictions of pre-big bang cosmology~\cite{Gasperini:1995dh}. As we have seen, it could also be one of the nice predictions of inflation in string theory. Even better than this, it might allow us to discriminate between different realizations of inflation in string theory.

\smallskip

{\bf \noindent Acknowledgements}

\smallskip

It is a pleasure to thank Kostantinos Dimopoulos, John Donoghue, Arthur Hebecker, Nemanja Kaloper, Marco Peloso, G\"unter Sigl, John Terning and Michele Trapletti for useful discussions.

\end{document}